\documentclass[12pt,letterpaper,]{article}
\usepackage{jheppub}

\usepackage{amsmath}
\usepackage{amsfonts}

\usepackage{graphics,epsfig}
\usepackage{graphicx, color}
\usepackage{slashed}
\usepackage{amsmath}
\usepackage{amssymb}
\usepackage{bbm}
\usepackage{epsfig}
\usepackage{yfonts}
\usepackage{graphicx}
\usepackage{caption}
\usepackage{subcaption}
\newcommand{\labell}[1]{\label{#1}}

\newcommand{\be}{\begin{equation}}
\newcommand{\ee}{\end{equation}}
\newcommand{\bea}{\begin{eqnarray}}
\newcommand{\eea}{\end{eqnarray}}
\newcommand{\ba}{\begin{eqnarray}}
\newcommand{\ea}{\end{eqnarray}}

\newcommand{\beq}{\begin{equation}}
\newcommand{\eeq}{\end{equation}}
\newcommand{\beqa}{\begin{eqnarray}}
\newcommand{\eeqa}{\end{eqnarray}}
\newcommand{\beqar}{\begin{eqnarray*}}
\newcommand{\eeqar}{\end{eqnarray*}}

\newcommand{\reef}[1]{(\ref{#1})}

\newcommand{\lp}{\ell_{\mt P}}

\newcommand{\ren}{R\'enyi\ }

\newcommand{\Tr}{{\rm Tr}}

\def\({\left(} \def\){\right)}
\def\[{\left[} \def\]{\right]}

\def\pd{\partial}
\def\=d{\, {\buildrel \rm def  \over =} \,}

\def\beq#1{\begin{equation} \label{#1}}
\def\ben{\begin{equation*}}
\def\een{\end{equation*}}
\def\bequa{\begin{eqnarray}}
\def\eequa{\end{eqnarray}}

\def\Tr{\mathop{\mathrm{Tr}}}

\def\beq{\begin{equation}}

\newcommand{\bseq}{\begin{subequations}}
\newcommand{\eseq}{\end{subequations}}

\renewcommand{\Re}{\mathop{\rm Re}\nolimits}

\def\HH{ \mathbb{H}}
\def\RR{ \mathbb{R}}

\newcommand{\mt}[1]{\textrm{\tiny #1}}

\newcommand{\rhoa}{\rho_\mt{A}}
\newcommand{\mue}{\mu}  %\bar{\mu}}  %_\mt{E}}
\newcommand{\norm}{{\cal N}_\mt{A}}

\newcommand{\taue}{\tau_\mt{E}}

\newcommand{\lstar}{\ell_*}

\arxivnumber{1407.nnnn [hep-th]}

\title{Charged R\'enyi entropies and holographic superconductors}

\author[a,b]{Alexandre Belin,}
\author[b,c]{Ling-Yan Hung,}
\author[a,b]{Alexander Maloney,}
\author[a]{Shunji Matsuura,}

\affiliation[a]{Departments of Physics and Mathematics, McGill
University, Montr\'eal, Qu\'ebec, Canada}
\affiliation[b]{Department of Physics, Harvard University,
Cambridge, MA 02138 USA}
\affiliation[c]{Department of Physics, Fudan University,
Shanghai, China}

\newcommand\adm[1]{{\color{magenta} ADM: #1}}

\emailAdd{alexandre.belin@mail.mcgill.ca}
\emailAdd{lhung@physics.harvard.edu}
\emailAdd{maloney@physics.mcgill.ca}
\emailAdd{matsuura@physics.mcgill.ca}

\abstract{Charged \ren entropies were recently introduced as a measure of entanglement between different charge sectors of a theory.  We investigate the phase structure of charged \ren entropies for CFTs with a light, charged scalar operator. The charged \ren entropies are calculated holographically via areas of charged hyperbolic black holes.  These black holes can become unstable to the formation of scalar hair at sufficiently low temperature; this is the holographic superconducting instability in hyperbolic space.  
This implies that the \ren entropies can be non-analytic in the \ren parameter $n$.
We find the onset of this instability as a function of the %we vary parameters of the CFT such as 
charge and dimension of the scalar operator. We also comment on the relation between the phase structure of these entropies and the phase structure of a holographic superconductor in flat space. }

\begin{document}

\maketitle

%%%%%%%%%%%%%%%%%%%%%%%%%%%%%%%%%%%%%%%%%%%%%%%%%%%%%%%%%%%%%%%
%%%%%%%%%%%%%%%%%%%%%%%%%%%%%%%%%%%%%%%%%%%%%%%%%%%%%%%%%%%%%%%
%%%%%%%%%%%%%%%%%%%%%%%%%%%%%%%%%%%%%%%%%%%%%%%%%%%%%%%%%%%%%%%

\section{Introduction}

%Quantum theory is essentially non-local; there are non-trivial correlations among subsystems of a quantum state.
%For a typical low energy state such as a ground state of quantum field theories, 
%the amount of information stored within a subsystem is quite small and most of the information is in correlations 
%among subsystems.
%This ignorance of an observer who is accessible to only
% a subsystem 
% is

%ground states of  quantum field theories, most of the information is stored 
% in correlations among subsystems.
%Compared to classical states, knowing states within a subsystem will not tell much about the total states.
% This ignorance is

%Hello everyone!  Today we're gonna show you a dangerous phase transition on a potato chip at Montreal temperature (${1/ 2\pi}$F or $-17^{\circ}$C).
%We expect that you are so sophisticated and don't even need any introduction, right? You help us save our time?? i know i know...

%Quantum information theory provides a useful framework to characterize 
The low energy states of a quantum field theory typically exhibit a high degree of spatial entanglement.
%the spatial entanglement structure of quantum field theory.
%For most of experimentally accessible states such as ground states or low energy states of quantum field theories, quantum information is stored within the correlations between different subsystems, rather than inside individual subsystems.
%Entanglement entropies have proven useful probes of the spatial entanglement structure of quantum field theory.
%
%This lack of information within each subsystem is
%quantified by  entanglement entropy and \ren entropy.
%For a given reduced density matrix 
To characterize this entanglement precisely, consider a quantum field theory in state $\rho$, with space divided into two parts $A\cup B$.
The \ren entropies $S_n$ are the moments of the reduced density matrix $\rhoa={\Tr}_B \rho$ for subsystem $A$
%, the \ren entropy $S_n$ with \ren parameter $n$ is defined by 
\bea
S_{n}={1\over 1-n}\log {\Tr} [\rhoa^n].
\label{def renyi}
\eea
The entanglement entropy $S_{EE}$ is 
$S_{EE}=\lim_{n\to1}S_{n}=-{\Tr} \rhoa\log \rhoa$.
These entropies encode the amount of information stored in correlations between $A$ and $B$, rather than in $A$ or $B$ separately.
%given by taking the $n \to 1$ limit:
%\bea
%S_{EE}=\lim_{n\to1}S_{n}=-{\Tr} \rhoa\log \rhoa,
%\label{def entangle}
%\eel
%which counts the number of states in the rest of the system that are consistent with all the measurements in
%the subsystem $A$. 
Entanglement entropies have played an important role in condensed matter physics \cite{wenx,cardy0},
quantum gravity \cite{Kabat:1994vj,rt0,mvr}, and quantum information \cite{Horodecki:2009zz}.

%There is a similar kind of entropy in thermodynamics; namely the thermal entropy $S_{therm}$.
%It is defined by (\ref{def entangle}) with $\rhoa$ replaced by the thermal density matrix.
%The thermal entropy satisfies the universal law of thermodynamics: $dF=-S_{therm}dT$ where $F$ and $T$ are the Helmholtz free energy and the temperature.
%In a system where particle number or some other charge $Q$ associated to a global symmetry is conserved, one can consider putting the system at finite charge density.  In that case, we have  $dF=-S_{therm}dT+\mu dQ$, where $\mu$ is a chemical potential and $Q$ is the number of particles.

In many cases we are interested in theories with a conserved charge $Q$ associated with a global symmetry, such as particle number.  When $Q$ is the integral of a local charge density one can ask how the entanglement depends on the distribution of charge between $A$ and $B$.
Very naively, one might expect that the entanglement between $A$ and $B$ should increase as charge is distributed more and more unequally between $A$ and $B$.  This is because one way of moving charge (say, particle number) into $A$ from $B$ is to create a particle-antiparticle pair, placing the particle in region $A$ and the anti-particle in region $B$; particle-antiparticle pairs created from the vacuum are naturally entangled, so this process should increase the entanglement entropy.  Of course, whether this naive expectation is true will depend on the details of the state and the theory.

To address this question we will follow 
%\adm{go through argument}
%these \ren entropies can be generalized to give more 
%
%A natural question is whether one can introduce 
%To address this, it is natural to introduce a notion of chemical potential in the entanglement entropy
%\footnote{The analogy between thermodynamics and quantum information was addressed recently in \cite{Bhattacharya:2012mi, Blanco:2013joa}}.
%This question was addressed in 
\cite{Belin:2013uta} and define the charged \ren entropy:
 \bea
S_{n}(\mue) = {1\over 1-n}\log \Tr \left[\,\rhoa\, \frac{e^{\mue\,
Q_\mt{A}}}{\norm(\mue)}\right]^n\,.
 \labell{charen}
 \eea
 The parameter $\mue$ is known as an entanglement chemical potential, and
 %The reduced density matrix 
% $\rhoa$ is the same as the one with zero entanglement chemical potential
%and 
$Q_\mt{A}$ measures the amount of charge in subsystem $A$; it is the integral over A of the local charge density. 
 %depends. 
 $\norm(\mue)\equiv \Tr [\rho_Ae^{\mue\,Q_\mt{A}}]$ is a normalization factor.
 Applications of  charged \ren entropies include the
supersymmetric \ren entropies \cite{Nishioka:2013haa}
and the characterization of symmetry protected topological phases \cite{ourSPT}.

We emphasize that $\mu$ is not the physical chemical potential of the system, but rather is a formal parameter used to weight the entanglement in different charge sectors.
Nevertheless, just as thermodynamic entropies can undergo phase transitions as temperatures and potentials are varied, the entanglement entropies 
$S_n(\mue)$ can undergo phase transitions as 
%As in the case of thermodynamical phase transitions, quantum information quantities can change discontinuously as we change parameters such as
the \ren parameter $n$ and the entanglement chemical potential $\mu$ are varied.
%Next question is whether there is a phase transition in the \ren entropy as one changes the parameters; the \ren parameter $n$ and the entanglement 
%chemical potential $\mue$.
These phase transitions reflect essentially non-analytic features of the spatial entanglement structure of quantum field theory ground states.

\ren phase transitions %in the absence of the entanglement chemical potential 
were investigated in \cite{Belin:2013dva} in the case $\mue=0$.
The authors considered 
large $N$ conformal field theories which are dual to semi-classical Einstein gravity in Anti-de Sitter space plus matter.
When the entangling surface $\partial A$ is a sphere, $S_n$ is related to the entropy of a black hole with hyperbolic event horizon.
%conformal field theories in $d$-dimensional space-time, and in particular  
%which are dual to Einstein gravity coupled to 
%with 
They showed that $S_n$ is non-analytic in $n$ if the dimension $\Delta$ of the lightest scalar operator ${\cal O}$ in the theory is sufficiently small.
%, which is dual to the .
 %$\mathcal{O}_{\Delta}$ of conformal dimension $\Delta$.   dual to Einstein gravity with a scalar. 
In particular, $\partial_n^2 S_n$ becomes discontinuous at some $n=n_c$, with $1 < n_c < \infty$, if
\bea
\frac{d-2}{2} <\Delta<\frac{d+\sqrt{d}}{2} \,. \labell{bound}
\eea
Here $d$ is the space-time dimensionality of the CFT and the first inequality is the unitarity bound.  The point is that the field $\phi$ dual to ${\cal O}$ will become unstable if the black hole temperature is sufficiently small. % operator 
When $n<n_c$, $S_n$ is computed by the entropy of a Einstein gravity black hole.  But when $n>n_c$ the Einstein black hole becomes unstable and the scalar operator gets a non-zero expectation value near the horizon.
In this phase the entangling surface hosts a localized "impurity" operator with non-vanishing expectation value.
We emphasize that this argument relies crucially on the fact that we are studying theories at large $N$.
%This situation is similar to the case of an impurity problem. 
%In general, 

Similar results have been found in purely QFT analysis  \cite{sachdevON,Steph-Rnyi-tran}.  
These authors consider
%showed that an entangling surface can host localized operators (impuritied).
%As one increases the \ren parameter $n$, 
the renormalization group flow of the coupling constants associated to impurity operators localized on the entangling surface. %changes. 
At a certain value of $n$, the beta functions for these couplings change sign and the system undergoes a phase transition.  For example, in 
\cite{sachdevON} it was argued that the \ren entropies of the $O(N)$ model are non-analytic at $n=7/4$.
%the localized operators on the entangling surface receive more quantum corrections. 
%
%at the entangling surface becomes very strong 
%we are essentially increasing the coupling strength of the defect, or  ``impurity'' localized at the entangling surface, 
%This quantum effect changes the renormalization group flow of the impurity coupling constants, and 
%eventually  the system undergoes a phase transition.
%This kind of phase transitions were observed in.
%Because of this phase transition, the \ren entropy is not an analytic function of the \ren parameter.

One consequence of this is that the replica trick -- where one computes $S_{EE}$ by computing $S_n$ at integer values of $n$ and analytically continuing to $n\to 1$ -- must be treated with care. %, as one cannot necessarily assume that the \ren entropies are analytic in $n$.
%Since the computation of the entanglement entropy (\ref{def entangle}) as a limit of the \ren entropy
%(\ref{def entangle}) relies on the analyticity of the \ren entropy near $n=1$, this phase transition underlines possible subtleties in using the replica trick. 
However, we emphasize that in the above analysis $n_c$ was shown to be always greater than 1, and approaches $1$ only as the dimension $\Delta$ approaches the unitarity bound (at which point the operator should decouple from the theory).
%
%. It decreases as $\Delta$ decreases and reaches $n_c=1$ exactly at the unitarity bound. 
Thus the \ren entropies are still analytic in a finite neighbourhood of $n=1$. %\footnote{Provided the spectrum is discrete.}
%. If analyticity was lost near $n_c=1$ however, the use of the replica trick would give misleading and incorrect results. 
So this result does not invalidate the derivation of the Ryu-Takayanagi formula by Maldacena and Lewkowycz \cite{Lewkowycz:2013nqa}, which  relied only on analyticity near $n=1$.
Further relations between the Ryu-Takayanagi formula and \ren entropies at $n\ne1$ were discussed in \cite{Headrick:2010zt, Fursaev:2012mp, Pastras:2014oka}. 
%In any case, the presence of these non-analiticities in $n$ for this particular example should be a sign of caution for the validitiy of the replica trick in more general settings.
%A second interesting consequence is that the min-entropy ($S_{\infty}$) -- which is the largest eigenvalue of $\rhoa$ -- will encodes the information of the conformal dimension $\Delta$ responsible for the phase transition.

In this paper we investigate the phase transitions of charged \ren entropies.  We consider large $N$ CFTs with global conserved charges that are holographically dual to Einstein-Maxwell-Scalar theory in AdS.
We will be interested in the charged \ren entropies of the ground state for spherical entangling surfaces; these \ren entropies can be computed by studying the CFT in hyperbolic space at finite temperature and charge density \cite{casini9, Belin:2013uta}.   In the bulk, the dual states are charged AdS black holes with hyperbolic event horizons \cite{renyi, Belin:2013uta}. 
%We will therefore study the phase transitions of charged, hyperbolic AdS black holes 
%Our task to explore the onset of phase transitions in charged \ren entropies therefore becomes one of exploring the stability of
%charged AdS black holes, an interesting and important question in Einstein gravity quite aside from its new found connections with quantum information. 

We will therefore study the phase structure of charged, hyperbolic black holes in AdS. %among charged scalar fields in these black hole backgrounds.  
In particular, we will consider instabilities where a charged scalar field -- dual to a CFT operator with global charge $q$ -- condenses near the horizon.
In this paper, we will find that charged entropies $S_n(\mu)$ will be non-analytic as we vary $n$ and $\mu$, provided the conformal dimension of the charged operator lies between
\be
\frac{d-2}{2} <\Delta_c(q)<\frac{1}{2}\left(d+\sqrt{d+\frac{8(d-2)q^2\mu^2}{8\pi^2R^2+(d-2)\mu^2}}\right) \,. \labell{chargedbound}
\ee
% It is therefore not surprising that 
%, and not surprisingly instabilities of the hyperbolic AdS black holes share various qualitative features with their flatly sliced counterparts, p
The setup is very similar to that used in the study of holographic superconductors  \cite{Hartnoll:2008kx,Gubser:2008px}, the only difference being that space is hyperbolic. %  Indeed, the phase transitions we are studying are precisely those of the holographic superconductor in hyperbolic space.   
The qualitative features of these instabilities are similar to those in flat space.
In particular, the high
temperature and large chemical potential behaviour  is identical to that of the flat-space holographic superconductor;
in these limits the curvature of the hyperbolic spatial slice is irrelevant. 
Thus, even though the charged \ren entropies under consideration probe only properties of the ground state, they contain information about the phase structure of physical thermodynamic quantities at finite temperature.  This echoes the results of \cite{haldane, fu}, that entanglement entropies of the ground state can be used to study phase transitions of the theory involving higher excited states.

In Section \ref{conformal mapping}, we review the relationship between the reduced density matrix for spherical entangling surfaces and 
thermal density matrices in hyperbolic space.
%, and 
%may be mapped to a thermal density matrix on hyperbolic space by a conformal mapping. 
%We 
In section \ref{bulk} we relate this to %how this is related to 
black hole entropy %charged \ren entropies 
using AdS/CFT, and describe the phase transition using both analytic and numerical techniques.  We compute the critical \ren parameter $n_c$ numerically using a shooting method. In Section \ref{discussion}, we discuss these results and relate them to the phase structure of 
the
%explain their relevance in understanding the phases of the 
holographic superconductor in certain limits. %We also comment on several interesting aspects of our results.

%\subsection{can we add more on:}
%\begin{itemize}
%\item entanglement entropy of finite density system
%\item thermal phase transition and quantum phase transition
%\item QFT computation of this phase transition.
%\end{itemize}

%We consider the generalized \ren entropies
% \ba
%S_{n}(\mue) = {1\over 1-n}\log \Tr \left[\,\rhoa\, \frac{e^{\mue\,
%Q_\mt{A}}}{\norm(\mue)}\right]^n\,.
% \labell{charen}
% \eea
%for the case of a CFT with a scalar operator $\mathcal{O}$ of dimension $\Delta$. 
%%%%%%%%%%%%%%%%%%%%%%%%%%%%%%%%%%%%%%%%%%%%%%%%%%%%%%%%%%%%%%%%%

\section{From entanglement entropy to thermal entropy \label{conformal mapping}}

%Let us begin by considering the reduced density matrix $\rho$ for half-space.
In a relativistic 
theory, an observer undergoing constant acceleration has causal access only to part of the space-time, known as the Rindler wedge, which is separated from the rest of the space-time by an event horizon.
%quantum theories, one can separate a subsystem that an observer can access from the rest of the system by boosting.
%A constantly boosted observer in Minkowski space can access only a quarter of the entire space-time (the Rindler wedge) which is separated from the
%rest of space-time by the Rindler horizon.
This Rindler wedge is the causal development of half-space, and a Rindler observer is in a thermal state due to the Unruh effect.
From this one concludes that the reduced density matrix associated with the half-space 
%In fact, one can show that the reduced density matrix of a half space 
is just the thermal density matrix on Rindler space \cite{Kabat:1994vj}.
%The Rindler space is the causal development of the half space:
%the set of points whose entire causal curves necessarily intersect the spherical region.
In a conformal theory, this can be generalized to relate the reduced density matrix for a spherical region and 
%Recalling that a plane and a sphere are related by conformal transformations,
%one can generalize this equivalence to one between the reduced density matrix of a spherical region and 
the thermal density matrix on a hyperbolic space %in conformal field theories
\cite{casini9, renyi}.  We will now review this argument.

%We begin by reviewing some results from %riefly review some results in 
%\cite{casini9, renyi}.% which are relevant to the current paper.
Consider a quantum field on a $d$-dimensional flat space in Euclidean signature
\bea
ds^2_{R^d} &=& dt_E^2 + dr^2+r^2 d\Omega_{d-2}^2
\label{flat coord}
\eea
where $t_E$ is the Euclidean time, $d\Omega_{d-2}$ is the volume element of a unit sphere, and
$r$ is the radial coordinate.
We are interested in the %computing the 
reduced density matrix of the $d-1$ dimensional ball of radius $R$ centered at the origin. %, taking , i.e. $(t_E=0,r<R)$.
%As mentioned, a sphere and a half space are related by conformal transformations and the reduced density matrix of a half space is
%equivalent to the thermal density matrix on Rindler space.
%Therefore, the reduced density matrix of a spherical region is equivalent to a thermal density matrix of $S^1\times H^{d-1}$, which is conformally
%equivalent to the Rindler space.
To compute this, we use the fact that flat space metric is conformal to $S^1\times \HH^{d-1}$:
%Consider the following conformal transformations
\bea
ds^2_{R^d} &=&  \Omega ^{-2}
ds^2_{S^1\times \HH^d} %\cr
 = \Omega ^{-2}
(d\tau_{E}^2+R^2(du^2+\sinh^2 u d\Omega^2_{d-2}))
\label{hyperbolic coord}
\eea
where
\bea
\Omega=\left|1+\cosh \left(u+i{\tau_E\over R} \right) \right|.
\eea
%are flat space coordinates $(t_E,r)$ in (\ref{flat coord}) 
and the $S^1\times \HH^{d-1}$ coordinates $(\tau_E,u)$ are defined by %in (\ref{hyperbolic coord}) are related by
\bea
\exp \left[ {-\left(u+i{\tau_E\over R}\right)} \right]={R-(r+it_{E})\over R+(r+it_{E})}.
\label{coords}
\eea
This conformal map has two important features.  First, the $\tau_E=0$ slice of the $S^1\times \HH^{d-1}$ geometry covers only the interior of the ball $(t_E=0, r=R)$; the entangling surface at $(t_E=0,r=R)$ has been mapped to the boundary of the hyperbolic space $u\to\infty$.
In Lorentzian signature (taking $\tau_E=i\tau$), this coordinate patch would cover only the interior of the Causal region associated with the ball $(t=0, r<R)$.
%, $0\le \tau_E < 2\pi R$, 
%The 
%The $\tau_E$ direction forms closed loops around the entangling surface, which shrink to zero size at the entangling surface.
%We note 
Second, from (\ref{coords}) we see that Euclidean time coordinate $\tau_E$ is periodic with period $2 \pi R$.
Putting this together, we see that the reduced density matrix is (up to a unitary transformation which does not enter into the trace)
% a finite temperature density matrix 
%Thus this is the Euclidean geometry which computes 
%In the hyperbolic conformal frame where the entangling surface has been pushed to spatial infinity,  $\tau$ has a fixed period related to the temperature. 
%This comes from the fact that the conformal factor $\Omega$ diverges as we approach the entangling surface.
%
%One can see that the angular direction surrounding the entangling surface $(t_E,r)=(0,R)$ is mapped to the time direction $\tau_E$
%on the hyperbolic space and the entangling surface, which is the fixed point of the rotational generator $-\pd/\pd \tau_E$,
%is mapped to the boundary of the hyperbolic space $u\to\infty$.
%
%The reduced density matrix can be expressed as
\bea
\rho={1\over Z_1}e^{-2\pi R H_E}
\label{density matrix}
\eea
Here  $H_E=i{\pd\over \pd \tau_E}$ is the Hamiltonian which generates translations in $\tau$ and $Z_1$ is a normalization factor.
Thus the reduced density matrix for a spherical region of radius $R$ is equivalent 
to the thermal density matrix on a hyperbolic space with radius $R$ and temperature 
$T_0=1/2\pi R$.
%Let us denote the thermal partition function on the hyperbolic space at the temperature $T=1/2\pi R n$ as
%$Z_n$.
From this, one can show that the \ren entropy (\ref{def renyi}) is %can be represented as
\bea
S_{n}={1\over n-1}(n\log Z_1-\log Z_n),
\eea
where $Z_n$ is the thermal partition function of the hyperbolic space at $T=T_0/ n$.
%Thus the \ren entropies with larger \ren parameter $n$ are associated with lower temperature %physics.

In a theory with a global conserved charge $Q$, one can generalize \reef{density matrix} to  
%the charged \ren entropy \ref{charen} is
%One can readily consider generalizations from the canonical ensemble above to the grand canonical ensemble as considered in \cite{Belin:2013uta}:
\bea
\rho_\mu%{\text{therm}} 
=\frac{e^{-H_E/T_0+\mue Q}}{Z(T_0,\mue)} \,.
 \labell{eq:rho0x}
\eea
where %$Q$ is the charge operator and 
$Z(T,\mue)$ is a partition function evaluated at temperature $T$ and chemical potential $\mue$.  %Following through the conformal transformation described above, one finds that $Q$ is simply the total amount of charge within region $A$. i.e.
The operator $Q$ measures the charge on hyperbolic space $\HH^{d-1}$, which -- from the conformal transformations above -- simply measures the amount of charge in region $A$.
%\be
%Q=Q_A.
%\ee
%Therefore, the thermal density matrix 
So \reef{eq:rho0x} is %precisely 
the generalized reduced density matrix introduced in \reef{charen}.

Note that $\mu$ can be interpreted as the time component of a background gauge field $B_\mu$
which couples to the charge density, via $B_{\tau_E}=\mue/2\pi R$ \footnote{We denote this background gauge field $B$ to avoid confusion with the bulk dynamical $A$ of the next section.}
 \beq
\mu=\oint B = \int_0^{2\pi R} B_{\taue} d\taue\,.
 \labell{round}
 \eeq
%Recall that in the flat conformal frame 
Since the $\tau_E$ circle shrinks to zero at the entangling surface,
%Therefore the c
this chemical potential introduces a %real 
magnetic flux localized at the entangling
surface \cite{Belin:2013uta}. 
%One can thus alternatively view this chemical potential as inserting some \emph{Dirac sheet}\footnote{as a generalization to Dirac strings} carrying a magnetic flux $-i\mu$ along the entangling surface. Note that the inclusion of this chemical potential is different from the reduced density operator of a finite density system:
%the charge operator $Q$ acts only on the subsystem inside the spherical region. 
%One naturally defines t
The charged \ren entropies
 \beq
S_{n}(\mue) = {1\over 1-n}\log \frac{Z(T_0/n,\mu)}{ Z(T_0,\mu)^n}.
 \labell{cow}
 \eeq
%These are novel measures of entanglement giving additional information 
measure the entanglement in the
(flat space, zero charge density) ground state of a theory, weighted by the charge contained in region A. 
%The charged \ren entropies are thus computing entanglement between different charged sectors of the theory.
%To compute the charged \ren entropies, we use t
In fact, using the
%Substituting the standard
thermodynamic identity 
%for the grand canonical ensemble thermal entropy
 \beq
S_{\text{therm}}(T,\mue)=-\left.{\pd F(T,\mu)\over \pd T}\right|_{\mu} ={\pd
\over \pd T}(T\log Z(T,\mue))\big|_{\mu}\,.
 \label{thermal S}
 \eeq
this can be written in terms of the standard grand canonical ensemble thermal entropy
$S_{therm}(T,\mu)$:
% into \reef{cow} we find the relation between the charged \ren entropy and grand canonical ensemble entropy 
 \beq
S_{n}(\mu)={n\over n-1}{1\over
T_{0}}\int_{T_{0}/{n}}^{T_{0}}S_{\text{therm}}(T,\mu)\,dT\, .
 \labell{Rnyi-therm}
 \eeq

Finally, it is important to mention the cut-off dependence of the \ren entropy.
The \ren entropy, as well as the entanglement entropy, are divergent quantities unless we set a UV cut-off near the entangling surface.
On the other hand, the thermal entropy on hyperbolic space is a divergent quantity as the volume of the hyperboloid is infinite; we must set an IR cut-off near the boundary of hyperbolic space\footnote{For holographic theories, the black hole entropy of a hyperbolic horizon is also a divergent quantity and requires an IR cut-off near the boundary of AdS space.}.
In fact, these divergences are identical: they can be mapped into one another by the conformal transformation (\ref{hyperbolic coord}). We refer the reader to \cite{casini9} for more details.

%In the following, we assume the all the quantities are properly regularized.

\section{Holographic computations\label{bulk}}

%In this paper, we want to study the phase diagram of the charged \ren entropy.
%We are interested CFTs that have a gravitational dual description. 
%In pure Einstein gravity, the gravitational dual is the well-known AdS black hole with hyperbolic event horizon.
 %\ren entropies are given by equation (\ref{Rnyi-therm}).
%The \ren parameter $n$ is related to the temperature on hyperbolic space $T=(2\pi R n)^{-1}$.
%The thermal entropy on a hyperbolic space is dual to the black hole entropy of hyperbolic black holes.
%This fact was used to compute the entanglement entropy and the \ren entropy holographically in \cite{casini9, renyi}.
% 
%As one increases $n$, the effective temperature of the system decreases.
%In the following section, we discuss the gravitational solutions for Einstein gravity.

In this section we calculate charged \ren entropies for large $N$ CFTs which are dual to semi-classical Einstein 
gravity coupled to matter in AdS space. In this case $S_{therm}(T,\mu)$ is the entropy of a black hole entropy in the bulk with hyperbolic event horizon  \cite{casini9, renyi}.
The global conserved current of the boundary theory acts as a source for a bulk gauge field. %, which we will take to be governed by a Maxwell.  
We will also assume the existence of a scalar operator in the boundary theory, with dimension $\Delta$ and charge $q$, which is dual to a charged scalar field in the bulk.  
%We will therefore take the bulk to be  Einstein gravity coupled to a Maxwell gauge field and a charged scalar field. 
%The charged \ren entropies are given by equation (\ref{thermal S}) where we integrate over black hole areas in the bulk. 
We are therefore looking for charged black holes with hyperbolic event horizons in Einstein-Maxwell-Scalar theory.
These solutions are dual to the grand canonical ensemble of the
boundary CFT on $\RR \times \HH^{d-1}$.  We first describe the Einstein-Maxwell solutions and give the \ren entropies when there is no scalar condensate. We then look for scalar instabilities for the Einstein-Maxwell black hole at the linearized level.  We perform an analytic analysis of the extremal black hole, and show that instabilities do occur in this case. % has an $AdS_2$ factor.
We then solve the Klein-Gordon equation for the scalar in the non-extremal case, using a numerical shooting method starting from the horizon. This allows us to determine the location of the phase transition in $S_n(\mu)$ as a function of $\Delta$ and $q$. 
We focus here on 3-dimensional CFT but a similar phase transition will occur in any dimension $d\geq3$.

\subsection{Neutral black holes and scalar instability}

Before discussing the charged \ren entropies and the related charged hyperbolic black holes, we review a few result concerning uncharged black holes. %will review some results for standard holographic \ren entropies and their phase structure. 
When the entangling chemical potential vanishes, the dual gravitational solutions are hyperbolic black holes
\begin{equation}
ds^2 = -f(r) \frac{L^2}{R^2}\, d\tau^2 +
\frac{dr^2}{f(r)} + r^2\, d\HH_{2}^2\,,
\end{equation}
where $d\HH_{2}^2 = du^2+\sinh^2\!u\, d\phi^2$ is the metric on
$\HH_{2}$ with unit curvature and
\begin{equation}
f(r) = \frac{r^2}{L^2}-1-\frac{M}{r}
\end{equation}
In the entanglement entropy limit $n\to1$, we recover the massless hyperbolic black hole, which is AdS$_4$ in Rindler coordinates.
%, in agreement with the Ryu and Takayanagi conjecture. 
To compute the \ren entropies, we integrate over a range of temperatures (\ref{Rnyi-therm}). As $n$ increases, the integral includes lower and lower temperatures. 

We now consider a scalar field in the bulk of negative mass-squared, which is dual to an operator of dimension $\Delta < 3$ in the boundary CFT.  In this case the black hole may become unstable at a certain temperature $T_c$ (or equivalently at a certain $n_c$) at which the black hole undergoes a second order phase transition.  This was shown in \cite{Belin:2013dva}, although in earlier work the authors of \cite{Dias:2010ma}
noted a similar instability for topological black holes with compact horizon (see also \cite{Martinez:2004nb}).
This effect can be understood as follows.  In the extremal limit these black holes have a $AdS_2\times \HH_{2}$ near horizon geometry. Scalar fields with masses below the effective Breitenlohner-Freedman bound for the near-horizon $AdS_2$  (suitably corrected for the $\HH_2$ factor) will become unstable at low temperatures. We emphasize that this happens for uncharged black holes; for AdS black holes with flat or spherical horizons, such instabilities occur only at finite chemical potential.
%In \cite{Belin:2013dva}, it was found that in the presence of a scalar field, non-compact hyperbolic black holes undergo a phase transition as well.
%As mentioned in the introduction, 
When the scalar field is below this bound the black hole becomes unstable
and will decay to a black hole solution with scalar hair. % when (\ref{bound}) is satisfied. 
The corresponding boundary  operator acquires a non-zero expectation value.
%The authors showed that if the field theory has an operator with dimension
%\bea
%\Delta<\frac{d+\sqrt{d}}{2} \,. \label{range}
%\eea
%the black hole undergoes a phase transition and acquires scalar hair. 

%\footnote{In quantum field theories, %this situation is quite similar to an impurity problem.
%When we divide a system into two regions to define the reduced density matrix, one has to set a boundary condition 
%at this entangling surface.
%As one changes the value of $n$, the renormalization group flow of operators localized on the entangling surface
%changes. When the effect of the impurity at the entangling surface gets strong, it can induce a phase transition.
%This was explicitly demonstrated perturbatively in the $O(N)$ model and a spin chain system in
%\cite{sachdevON,Steph-Rnyi-tran}.}.
%Our goal in this paper is to explore the phases of the charged \ren entropy in the presence of scalar fields.
%We also address a previously undressed issue in neutral scalar fields.
%In AdS/CFT, we need to set boundary conditions for bulk fields.
%If the masses $m^2$ of the bulk fields are $m^2_{BF}+1\le m^2$, only Dirichlet boundary conditions are allowed.
%However, if the masses take values in $m_{BF}^2<m^2<m^2_{BF}+1$, more general boundary conditions are allowed \cite{Klebanov}.
%As was observed in \cite{Dias:2010ma}, the critical \ren parameter $n_{cr}$ can be close to 1 with Neumann boundary condition, or
%in double trace deformed CFT.
%We study the detail of the critical \ren parameter with Neumann boundary condition with or without entanglement chemical potential.
%

This instability implies that $S_n$ has a phase transition as a function of $n$. 
The \ren entropies are obtained from the thermal entropies by integrating once so $\partial_n ^2S_n$ will become discontinuous at some critical \ren parameter $n_c$.
In order to determine  the precise value of $n_c$ at which this transition occurs, it is necessary to study numerically the scalar wave equation in the black hole background. 
The results of \cite{Belin:2013dva} (see also \cite{sachdevON,Steph-Rnyi-tran}) show that $n_c>1$ as long $\Delta$ is above the unitarity bound.
In particular, if we compute the entanglement entropy by taking $n\to 1$ then $S_n$ is given by the Einstein black hole with vanishing scalar field.
We will see that this is no longer the case for charged \ren entropies: when $\mu\ne0$, it is possible for $n_c$ to be less than one.

%There is another important issue that was not addressed in detail in \cite{Belin:2013dva}.
%
%Our goal is to investigate the phase diagram
%
%
%
%
%
%one can use a non-standard boundary condition.
%double trace deformation.
%What is the critical value of $n$ for Neumann boundary condition? \cite{Dias:2010ma} 

\subsection{Charged black hole}

The Einstein-Maxwell-Scalar action with a negative
cosmological constant is\footnote{The scale $\lstar$ 
%appearing in the prefactor of the Maxwell term should be fixed by the 
depends on the details of the %boundary 
theory. With
this notation, the 4-dimensional gauge coupling is
$g_4^{\,2}=2\lp^{2}/\lstar^2$.
%\adm{do you mean $g_4$ here?}
}
\begin{equation}
I_{E-M} = \frac{1}{2\lp^{2}} \int d^{4}x\,\sqrt{-g}\left(\frac{6}{L^2} + \mathcal{R}
-\frac{\lstar^2}{4} F_{\mu\nu}F^{\mu\nu}-V(|\phi|)-\frac{1}{2}|\nabla \phi -iqA\phi|^2\right)\,.
\labell{action}
\end{equation}
We will take the potential to be a mass term $V(|\phi|)=\frac{1}{2}m^2|\phi|^2$ which, together with boundary conditions, fixes the conformal dimension $\Delta$ of the dual CFT operator.  We first consider charged hyperbolic black hole solutions with vanishing scalar field.
%, which are charged topological black hole. Their metric is given by
The metric is
\begin{equation}
ds^2 = -f(r) \frac{L^2}{R^2}\, d\tau^2 +
\frac{dr^2}{f(r)} + r^2\, d\HH_{2}^2\,,
\labell{bhmetric}
\end{equation}
with
\begin{equation}
f(r) = \frac{r^2}{L^2}-1-\frac{M}{r}+\frac{\rho^2}{r^{2}}
\labell{funct}
\end{equation}
The time coordinate is normalized 
 so that the boundary metric naturally is flat space in Milne coordinates: $ds^2_{CFT} =
-d\tau^2+R^2 d\HH_{2}^2$ \cite{casini9}. The bulk gauge field is
\begin{equation}
A = \left(\frac{2L\,\rho}{R\lstar\, r}-\frac{\mue}{2\pi R}
\right)d\tau\,.
\labell{gauge}
\end{equation}
%The chemical potential $\mue$ is fixed by requiring that the 
We will chose our gauge so that $A$ vanishes 
at the horizon  $r=r_H$, so that
\begin{equation}
\mue = 4\pi\frac{L\,\rho}{\lstar r_H} \,.
\end{equation}
The mass parameter $M$ is related to the horizon radius $r_H$ by
\begin{equation}
M = \frac{r_H}{L^2}(r_H^2-L^2)+\frac{\rho^2}{r_H}  \,.\labell{mass parameter}
\end{equation}
%Hence, we may 
We can rewrite $f(r)$ %(\ref{funct}) 
in terms of the horizon
radius $r_H$ and the chemical potential $\mu$ as
\begin{equation}
f(r) = \frac{(r-r_H)(16\pi^2r(r^2-L^2+r r_H +r_H^2)-r_H (\ell_*\mu)^2)}{16\pi^2L^2r^2} \,.
\end{equation}
The temperature of this black hole is %given by
\begin{equation}
%T = T_0\left( \frac{L}{r_H} + \frac{r_H^2}{2L} f'(r_H) \right)
T = \frac{T_0}2\, L f'(r_H)=\frac{T_0}2\, \left[
3\,\frac{r_H}{L}-\frac{L}{r_H}\left(1+\left(
\frac{\mu\,\lstar}{4\pi L}\right)^2\right)\right]
\labell{gambit}
\end{equation}
where $T_0=1/2\pi$ %is the temperature 
and prime denotes
differentiation with respect to $r$. The thermal entropy of the black hole is given by the
Bekenstein-Hawking formula: % and reads
\begin{equation}
S_{therm} = \frac{2 \pi}{\lp^{2}}  V_\Sigma\,r_H^{2}\,, \labell{thermal}
\end{equation}
where $V_\Sigma$ denotes the (appropriately regulated) volume of $\HH^{2}$.  As noted above, the large-volume divergence of $V_\Sigma$ is related to a UV divergence in the boundary theory \cite{renyi}.  In particular, $V_\Sigma$ 
can be regarded as a %Recall that tis a
function of $R/\delta$, the ratio of the radius of the entangling sphere to the
short-distance cut-off in the boundary theory.  The leading term is
%Further the leading contribution takes the form
 \be
V_\Sigma \simeq
2\pi\,\frac{R}{\delta}+\cdots\,,
 \labell{led99}
 \ee
Hence the corresponding \ren entropies begin
with an area law contribution. When there is no scalar condensate, the charged \ren entropies can be computed from (\ref{thermal}) and (\ref{Rnyi-therm}):
 \beqa
S_{n}(\mue) &=&   \pi V_\Sigma \left(\frac{L}{\lp}\right)^{2}\frac{n}{n-1}
\left[ \left(1+\frac{1}{4}\left(\frac{\mue \lstar}{2\pi L}\right)^2\right )
(x_1-x_n) + x_1^3-x_n^3\right] \nonumber\\ && \labell{renyi entropy}
 \eeqa
with
\begin{equation}
x_n = \frac{1}{3n} +
\sqrt{\frac{1}{9n^2}+ \frac{1} 3+\frac{1}{12}
\left(\frac{\mue \lstar}{2\pi L}\right)^2}\,.
 \labell{joke}
\end{equation}

%%%%%%%%%%%%%%%%%%%%%%%%%%%%%%%%%%%%%%%%%%%%%%%%%%%%%%%%%%%%%
\subsection{Scalar instabilities}

The Einstein-Maxwell black holes described above may become unstable at sufficiently low temperature in the presence of a scalar field. 
We will find the onset of the instability  by solving the wave equation for the scalar on the black hole background.
The endpoint of the instability will be a hairy black hole with a non-zero scalar field in the vicinity of the horizon.  
Thus there will be more than one classical gravitational solution that satisfies the same boundary conditions.
We should then compare free energies of these saddle point configurations and determine which one is thermodynamically preferable.  In general,  we expect that the dynamically stable saddle point is thermodynamically preferred \cite{GubserMitra}; indeed, in the holographic superconductor \cite{Hartnoll:2008kx} and the constant mode analysis \cite{Belin:2013dva}  scalar condensate phases were found to be thermodynamically preferable.  We expect that in the present case the dynamical instability analysis will agree with the thermodynamical analysis as well.

We note that the instability involves a scalar field which is not constant on $\HH_2$, as the constant mode on $\HH_2$ is non-normalizable.  Thus the hairy black hole will not have the full hyperbolic symmetry of the Einstein-Maxwell black hole, making its explicit construction a technically difficult task.  We will therefore leave the construction of hairy black hole solutions to future work; in the present paper we simply demonstrate that the black holes are unstable and find the onset of the instability.

To find the instability, we must study the Klein-Gordon equation for a charged (complex) scalar $\Phi$ on the black hole background:
\be
\left((\nabla^\mu-iqA^\mu)(\nabla_\mu-iqA_\mu) -m^2\right)\Phi=0 \,.
\ee
It is convenient to decompose the field in eigenfunctions of the Laplacian on $\HH_2$:
\be
\Phi =\frac{\phi(r)e^{\omega \tau}Y(\sigma)}{r}
\ee
where $\nabla_{\HH_2}^2 Y=-\lambda Y$.  For a normalizable mode on $\HH_2$, $\lambda>1/4$. 
The wave equation reduces to: % Schrodinger form:
\bea
\left(-\left(f(r)\frac{d}{dr}\right)^2+V(r)\right)\phi(r)&=0%-\omega^2\phi(r)\\
\label{schr}
\eea
with
\be
V(r)=\frac{f(r)}{r^2}\left(\lambda+rf'(r)+r^2m^2\right)+\left(\omega+iqA_\tau\right)^2 
\label{potential}
\ee
It is convenient to put equation (\ref{schr}) in Schrodinger form, by defining the tortoise coordinates $r^*$ by $dr^*=dr/f(r)$,
so that 
\bea
\left(-\left(\frac{d}{dr^*}\right)^2+V(r^*)\right)\phi(r^*)&=0%-\omega^2\phi(r)\\
\label{schr}
\eea

%The extra factor of $1/r$ is such that the wave equation resembles that of a Schrodinger problem, namely
An unstable mode corresponds to a solution of this equation with $\Re \omega>0$.
We note that when $q=0$, (\ref{schr}) is just a one dimension Schrodinger equation with potential $V(r)$, although in general $V(r)$ will be complex. %When $q\neq0$, this is no longer the case because the potential $V(r)$ is complex.  
%The Hamiltonian being no longer hermitian, the eigenvalues $\omega$ are typically complex and an unstable mode has . 
%However, as we are just interested in finding the onset
%of the instability.
We are interested in finding the onset of this classical instability, i.e. we seek a solution with $\Re(\omega)=0$.

%As explained in \cite{Dias:2010ma}, the onset of instability is actually given by $\omega=0$ in the gauge where $A_\tau(r_H)=0$. To see this, note that we can always remove any imaginary part of $\omega$ by a constant gauge transformation of $A_\tau$. Near the horizon, the effect of such a gauge transformation will simply be a negative constant shift in the potential (\ref{potential}) and the Klein-Gordon equation will be solved with $\Re(\omega)> 0$. Thus the onset of instability corresponds to $\omega=0$ in which case the potential is real and the problem reduces to a Schrodinger equation in tortoise coordinates.

Before performing the numerical analysis of the black hole stability, we can understand analytically when the black hole should become unstable. In the zero temperature limit, the black hole (\ref{bhmetric}) has an 
$AdS_2\times \HH_{2}$ near horizon geometry.  The AdS$_2$ and $\HH_2$ radii are 
\begin{equation}
L^2_{AdS_2}=\frac{2L^2_{AdS_{4}}}{f''(r_{ext})} \ \ \ \ \ \ \ \ L^2_{\mathbb{H}_{2}}=r_{ext}^2
\end{equation}
where $r_{ext}$ is the horizon radius of the extremal black hole. In addition, there is a constant electric flux on the near horizon $AdS_2$: 
\begin{equation}
F =  -\frac{\mu}{2\pi R r_{ext}} dvol_{AdS_2}
\end{equation}
In the near horizon region, one can therefore approximate the wave equation (\ref{schr}) by
\begin{equation}
\left[-\(\frac{\epsilon^2f''(r_{ext})}{2}\partial_\epsilon\)^2+\epsilon^2\frac{f''(r_{ext})}{2r_{ext}^2}(\lambda+r_{ext}^2m^2)+\(\omega-i\epsilon\frac{q\mu}{2\pi R r_{ext}}\)^2\right]\phi(\epsilon)=0 
\label{nh}
\end{equation}
where $\epsilon$ is the near-horizon radial coordinate $r=r_{ext}+\epsilon$.  The solutions to (\ref{nh}) can be found analytically, and one can then determine exactly when an instability occurs.  In fact, it is not necessary to even solve (\ref{nh}) explicitly to see the instability.  We can simply note that near the asymptotic boundary of the near-horizon $AdS_2\times \HH_2$, solutions to (\ref{nh}) will behave as $\phi(\epsilon) \sim \epsilon^{-\Delta_{eff}}$, where 
\be
\Delta_{eff}=1/2+\sqrt{m_{eff}^2L_{AdS_2}^2+1/2},~~~~~
m_{eff}^2=\frac{f''(r_{ext})}{2r_{ext}^2}(\lambda+r_{ext}^2m^2)-\frac{q^2\mu^2}{4\pi^2R^2r_{ext}^2} \,.
\ee
Thus $\phi$ behaves like a field of mass-squared $m_{eff}^2$ in the near-horizon AdS$_2$.  
%Note that 
% This could be checked directly from the ODE but one we can also argue that in the usual AdS/CFT language, complex conformal dimensions would correspond to time evolution with a complex frequency, and thus to a growth in time.
%by\footnote{There is also a term of the form $iqA^\mu\partial_\mu$ that we neglect in the following approximation. This term gives the term linear in $\omega$ in the potential (\ref{potential}) which we argued can be neglected when searching for the onset of instability.}
%\begin{equation}
%\left((\nabla^\mu-iqA^\mu)(\nabla_\mu-iqA_\mu) -m^2\right)\Phi\approx \left(\nabla_{AdS_2}^2 + \nabla^2_{\HH_2} - q^2 A^\mu A_\mu - m^2 \right)\Phi=0
%\end{equation}
%If we expand in eigenfunctions of $\nabla_{\HH_2}^2$, then $\nabla^2_{\HH_2} - q^2 A^\mu A_\mu$ just gives an overall constant addition to the effective mass-squared of the scalar field.
The black hole will become unstable when the effective mass-squared falls below the $AdS_2$ Breitenlohner-Freedman bound, 
$m_{eff}^2L_{AdS_2}^2<-1/4$, i.e. when $\Delta_{eff}$ becomes complex.

%Putting this together, w
We conclude that an instability will occur when
\be
% \notag \\
m^2L^2_{AdS_{4}}<-\frac{f''(r_{ext})}{8}-\frac{1}{4r_{ext}^2}+\frac{2q^2\mu^2}{4\pi^2R^2r_{ext}^2f''(r_{ext})} \label{deltacrit} \,.
\ee
The first term on the right hand side is the naive $AdS_2$ BF bound.  The second term is a correction term coming from the fact that the lowest eigenvalue of a normalizable mode on $\HH_{2}$ has $\lambda=1/4$; this effect makes the scalar more stable.  The final term is a correction term coming from the charge coupling $q^2A_\mu A^\mu$; this effect makes the scalar more unstable. 
%  We note that the scalar profile on $\HH_{2}$ makes the scalar more stable but the effect of charge is the opposite. 
Using the form of the black hole metric, we expect instabilities when
\be
-\frac{9}{4}<
m^2L^2_{AdS_4}<-\frac{3}{2}+\frac{2q^2\mu^2}{8\pi^2R^2+\mu^2}
\ee
The first inequality is the usual BF bound in AdS$_4$.

It is straightforward to generalize this to arbitrary dimension (we will omit the details for the sake of brevity).  In general, we find an instability when
\be
-\frac{d^2}{4}<m^2L_{AdS_{d+1}}^2<-\frac{d(d-1)}{4}+\frac{2(d-2)q^2\mu^2}{8\pi^2R^2+(d-2)\mu^2} \label{deltacrit} \,.
\ee
In terms of conformal dimension of dual operators, this gives
\be
\Delta<\frac{1}{2}\left(d+\sqrt{d+\frac{8(d-2)q^2\mu^2}{8\pi^2R^2+(d-2)\mu^2}}\right)\equiv \Delta_c(q) \label{deltabound}
\ee

\begin{figure}[htb]
\begin{center}
\includegraphics[width=1\textwidth]{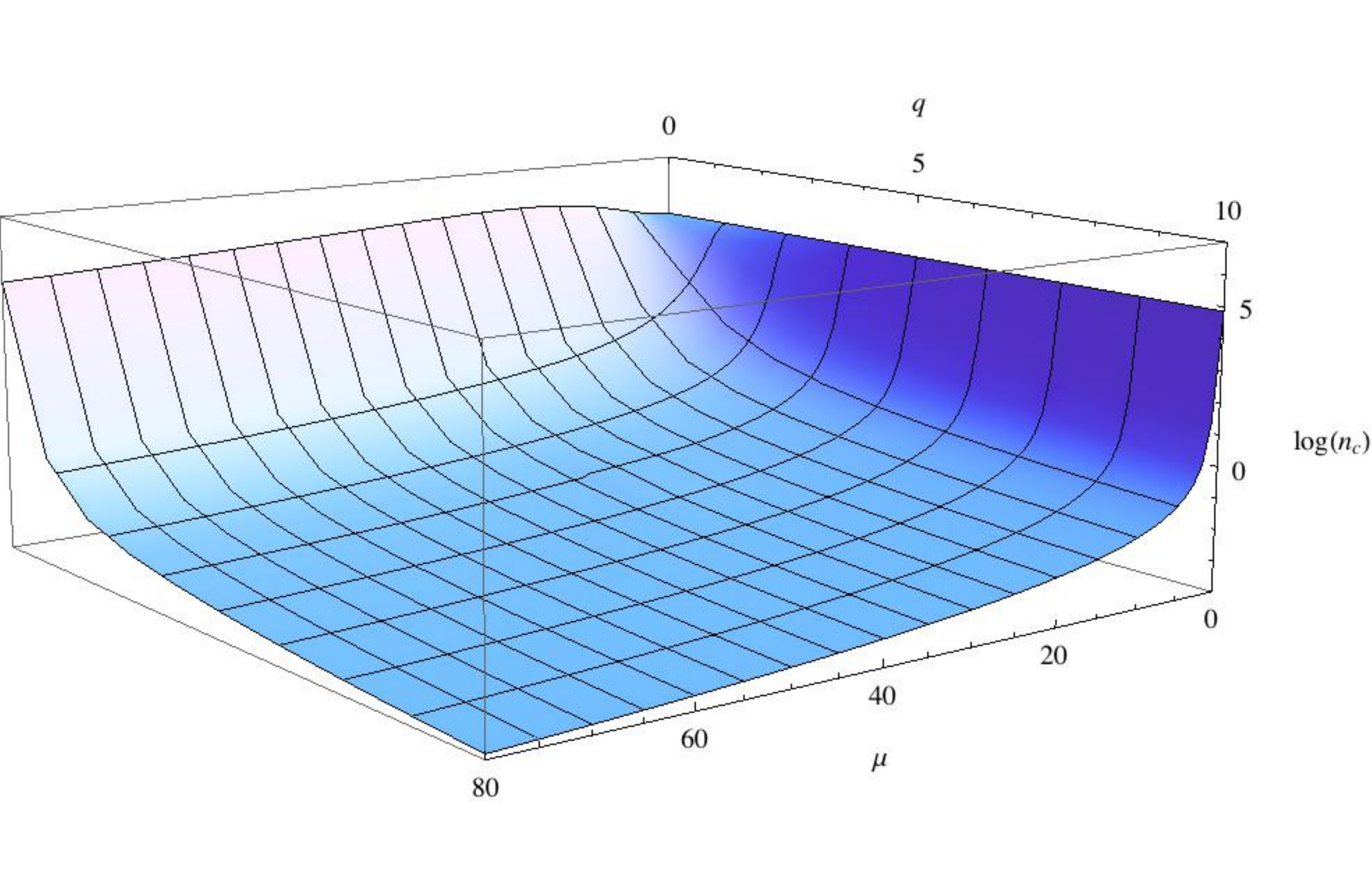}
\caption{$\log n_c$ as a function of $\mu$ and $q$ for $\Delta=2$. Every configuration above this surface is unstable.
}
\label{fixedDelta}
\end{center}
\end{figure}

We will now study the stability of the charged hyperbolic black holes by numerically solving the scalar wave equation\footnote{For the remainder of this section we set $L=R=\ell_*=1$}. Normalizability of the modes requires the field to be regular at the horizon; we can expand perturbatively for the field  near the horizon. We then use a shooting method to obtain desired boundary conditions near the boundary of $AdS_4$, namely
\be
\Phi(r)|_{r\to \infty}\propto r^{-\Delta}
\ee
In Fig. \ref{fixedDelta}, we show the results.  We show the critical value of the \ren parameter $n_c$ at which the phase transition occurs, as a function of $\mu$ and $q$ for $\Delta=2$.  We see that increasing the charge makes the configuration less stable for any $\mu\neq0$. For large enough $q$, increasing $\mu$ also renders the black hole less stable. However, for a neutral scalar, as we increase $\mu$ we make the black hole first more stable until some maximum value after which $n_c$ decreases again. 
We will return to discuss this non-monotonic behaviour in the next section.

\begin{figure}[htb]
\begin{center}
\includegraphics[width=1\textwidth]{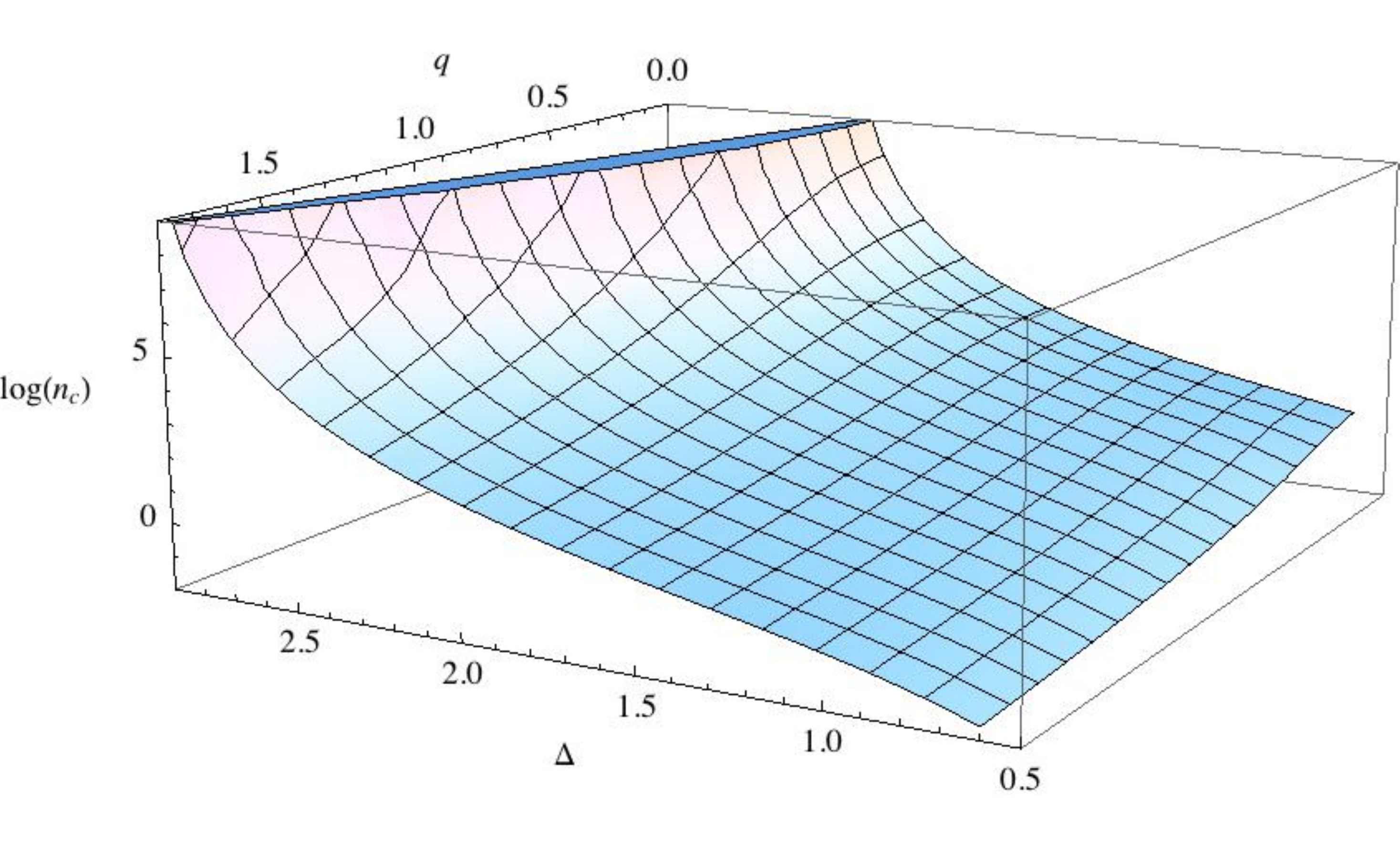}
\caption{The graph of $\log n_c$ as we vary $q$ and $\Delta$ for $\mu=5$. Every configuration above this surface is unstable.
}
\label{fixedmu}
\end{center}
\end{figure}

\begin{figure}[htb]
\begin{center}
\includegraphics[width=0.8\textwidth]{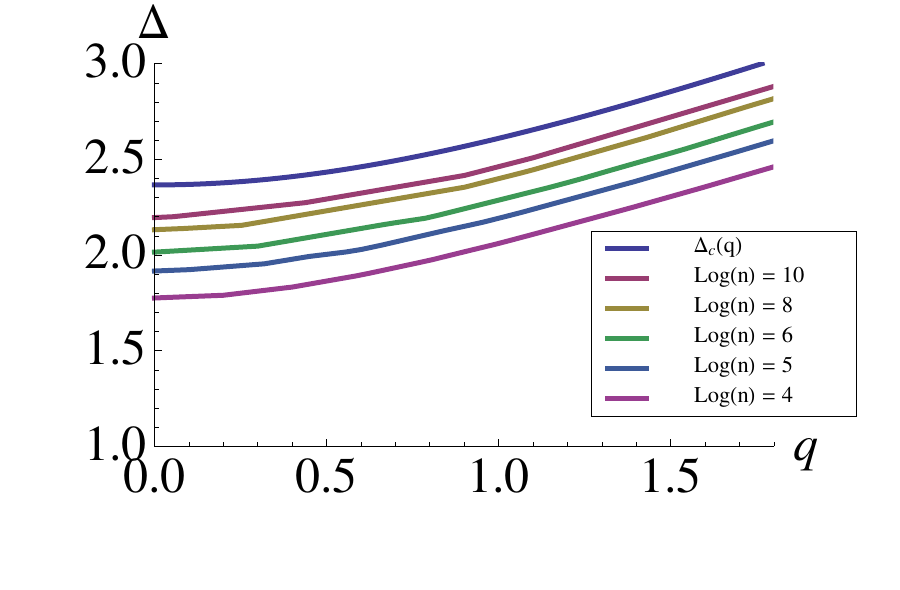}
\caption{The graph of $\Delta$ as we vary $q$ for $\mu=5$ and different values of $\log n$. As we increase $n$, we get closer and closer to the analytical estimate $\Delta_c(q)$.
}
\label{Delta_c}
\end{center}
\end{figure}

In Fig. \ref{fixedmu}, we show the value of $n_c$ for $\mu=5$, varying $\Delta$ and $q$. 
%A given CFT comes with its values of $\Delta$ and $q$ so we should think of this graph as the critical \ren parameter for $\mu=5$ as we move through the space of theories.  
As expected, we see that decreasing $\Delta$ or increasing $q$ makes the configuration less stable. From this graph, we can determine the critical value $\Delta_c$ (for a given $q$) at which the instability kicks in. 
%However, getting a precise value would require solving the differential equation close to zero temperature which is technically difficult. 
In Fig. \ref{Delta_c}, we plot the curves $\Delta(q)$ for different values of $n_c$ and show that the curves approach the analytic value derived in (\ref{deltacrit}) as $n_c\to\infty$. This confirms the effective mass analysis given above. % for charged hyperbolic black holes. 
%We leave the proof of this statement for future work.

\begin{figure}[htb]
\begin{center}
\includegraphics[width=1\textwidth]{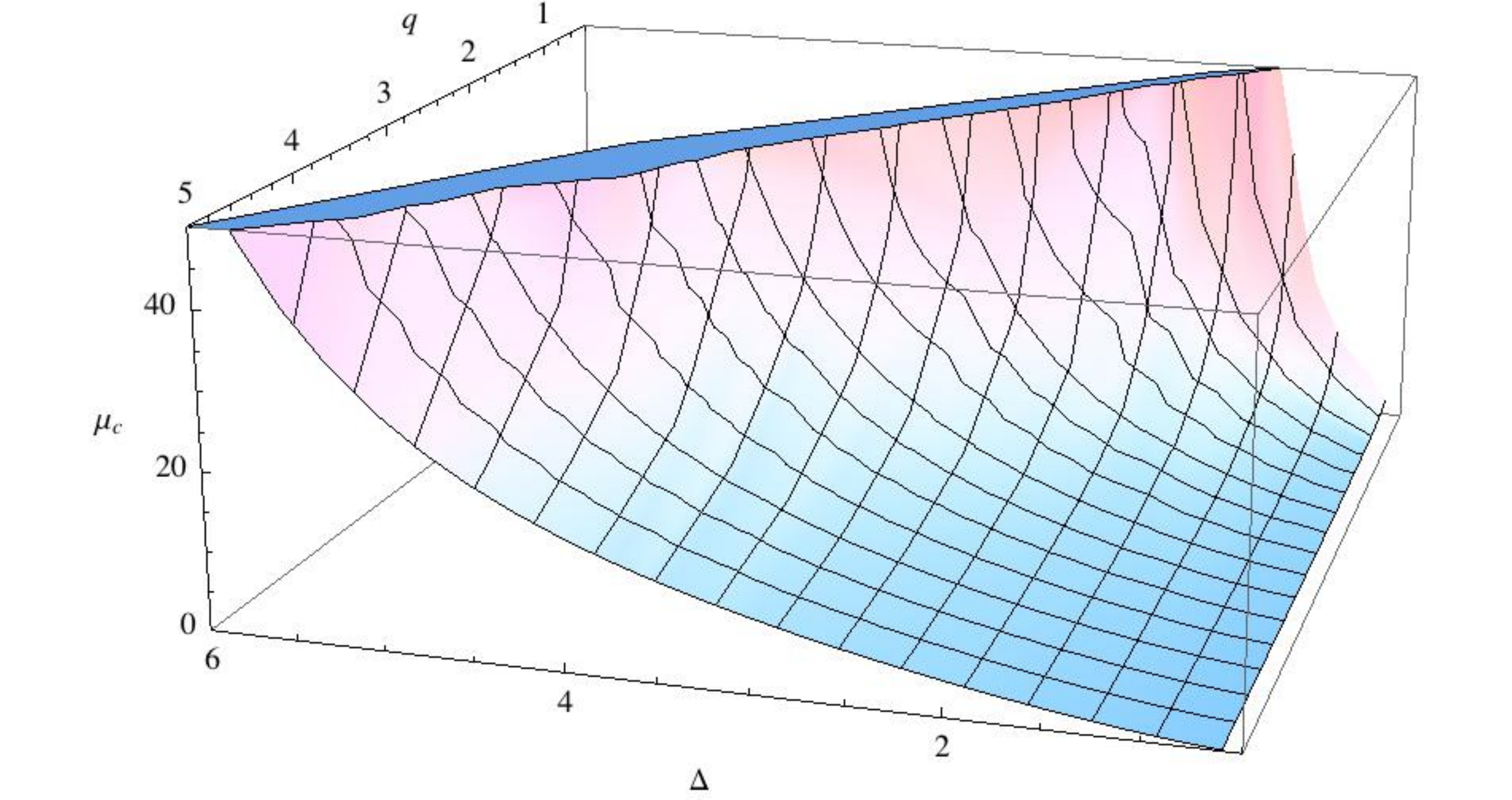}
\caption{The graph of $\mu_c$ as we vary $q$ and $\Delta$ at $n_c=1$. Every configuration above this surface is unstable.
}
\label{n=1}
\end{center}
\end{figure}

In Fig. \ref{n=1}, we show the critical chemical potential $\mu_c$ as a function of ($\Delta$,$q$) with $n_c=1$.  We see that, even when $n=1$, it is possible for the Einstein-Maxwell black hole to be unstable.  When $\mu>\mu_c$, the scalar condenses even though the only remaining defect inserted at the entangling surface is a Wilson line.  As we increase $q$ and/or decrease $\Delta$, $\mu_c$ decreases.  We note, however, that $\mu=0$ is always stable as long as $\Delta$ is above the unitarity bound, $\Delta > 1/2$.

\section{Discussion \label{discussion}}

\begin{figure}[htb]
\begin{center}
\includegraphics[width=1\textwidth]{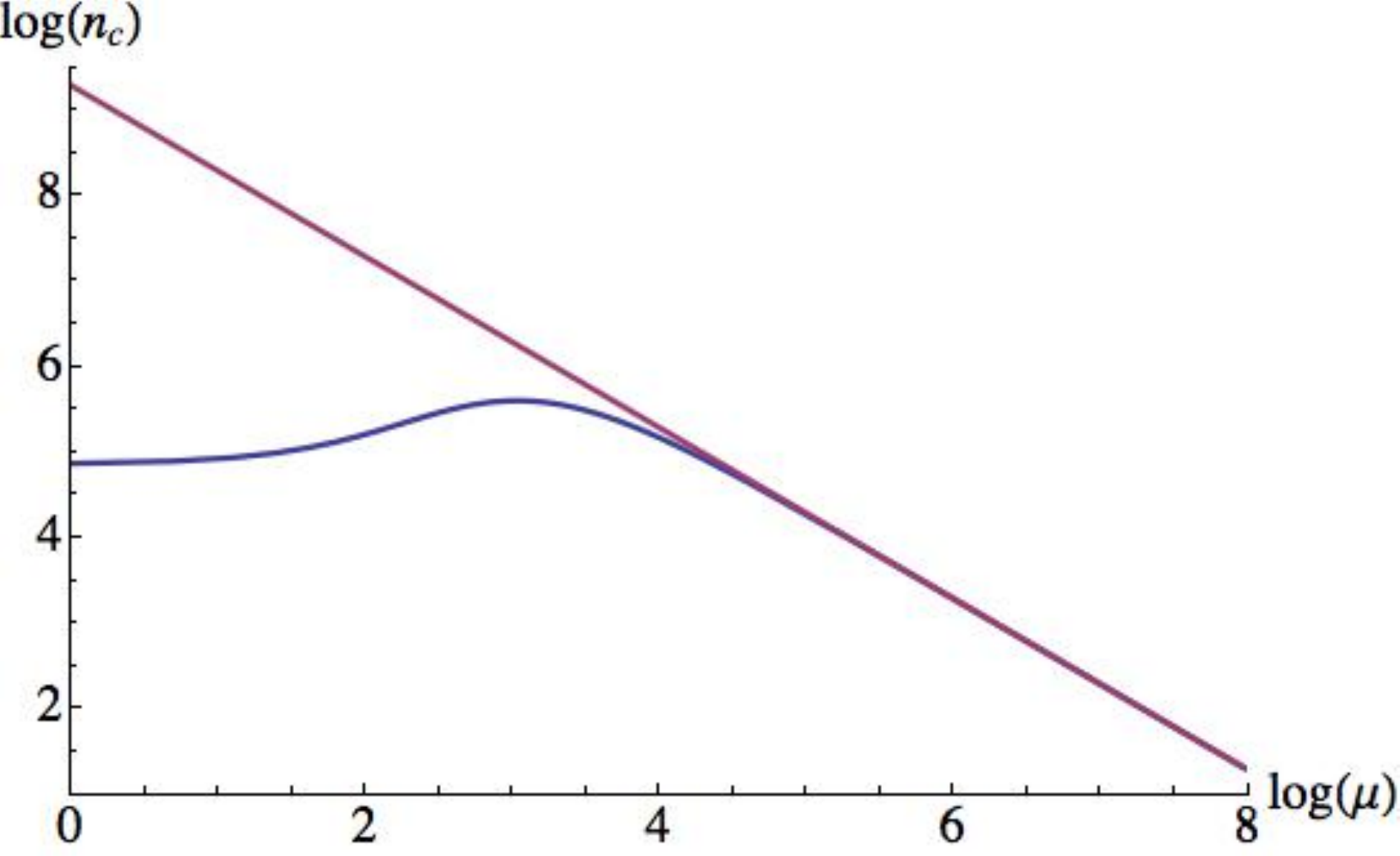}
\caption{The graph of $\log n_c$ as we vary $\mu$ for $q=0$ and $\Delta=2$. The blue line represents the critical \ren parameter and the red line represents the critical temperature for the field theory on flat space with physical chemical potential $\mu$.
}
\label{q=0}
\end{center}
\end{figure}

We have shown that the hyperbolic charged black hole can become unstable, and investigated the onset of the instability 
%for various bulk configurations %which translate into a discontinuity of charged \ren entropies  
as we vary $n$, $\mu$ and $\Delta$.  
We now comment on our results.  First, we note that for an uncharged scalar ($q=0$), the black hole can be unstable for sufficiently small  scalar mass. This is not surprising; for $\mu\ne 0$ this was already noticed in the holographic superconductor \cite{Hartnoll:2008kx} for  black holes with flat horizons. However, we have seen  that for hyperbolic black holes increasing the chemical potential first renders the solution {\it more} stable until we reach a peak of stability. Beyond that point, increasing $\mu$ renders the black holes more unstable as can be seen in Fig. \ref{q=0}. This is in {contrast} with flat  black holes, where no condensation can occur for $\mu=0$.
%; in that case, increasing $\mu$ only make the black hole more unstable.

In addition to the numerical method used above, one can also use a WKB method to determine when instabilities exist and to approximate the unstable modes. 
We will study this in the $q=0$ case, and use the WKB analysis to confirm the surprising behaviour described above.
%The wave equation for the scalar can be seen as a Schrodinger problem and we can use the WKB approximation to estimate the number of bound states of the potential.  
The number of bound states of the Schrodinger potential $V(r^*) $ can be estimated using the WKB integral:
\be
N_{bound~ states}=\int\sqrt{-V(r^*)}dr^*=\int_{rh}^{r0} \frac{\sqrt{-V(r)}}{f(r)}dr
\ee
where we integrate from the horizon $r=r_h$ up to the zero of the potential $V(r_0)=0$, with $r_0>r_h$.  The instability appears when $N_{bound~ states}\sim 1$.
Keeping the temperature fixed and increasing $\mu$, we find $N_{bound~states}$ first decreases and then increases again as we increase $\mu$. We plot of $\int\sqrt{-V(r^*)}dr^*$ against $\mu$ at $\Delta=2$ at 10 different temperatures in Fig. \ref{bump} below. One can see that for a given $n$ it decreases with $\mu$
before increasing again for sufficiently large $\mu$, signifying that the system is more stable as $\mu$ increases at small $\mu$, but the trend is reversed for larger values of $\mu$. 
%This despite the fact that the WKB approximation is not expected to be good in the absence of any perturbative parameter. 
As a note of caution however, we emphasize that -- in the absence of any small perturbative parameter -- this WKB approximation should at best be taken with a grain of salt, although it does reproduce the qualitative features of the numerics.
%We note that for $m^2 < -2$, the effective potential $V(r)$ becomes unbounded below as $r\to\infty$ \cite{Belin:2013dva}, and this is independent of the charge $q$ or $n$ and $\mu$. One can readily check that  $\int_{rh}^{r0} \frac{\sqrt{-V(r)}}{f(r)}dr$ diverges logarithmically with large $r$.  And yet, as is demonstrated by our numerical solution, the black hole remains stable above some critical temperature or below some chemical potential $\mu$.

\begin{figure}[htb]
\begin{center}
\includegraphics[width=1\textwidth]{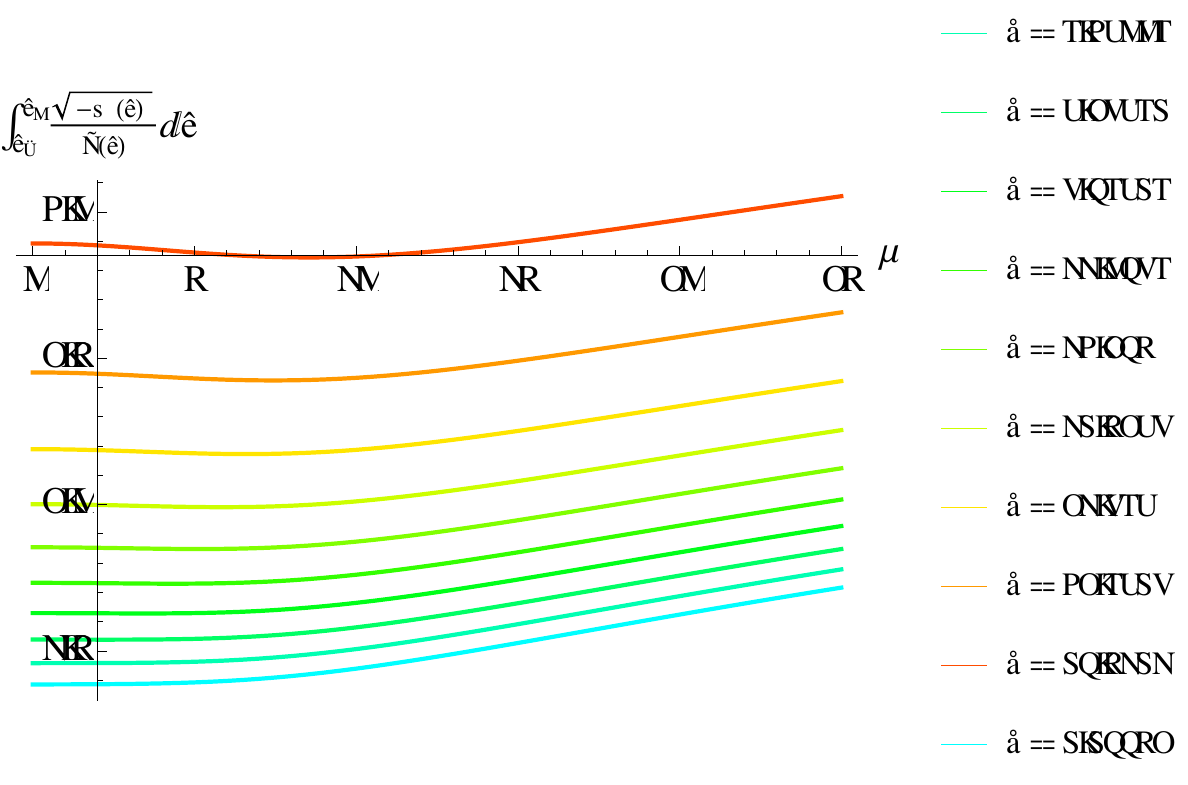}
\caption{Plots of $\int_{r_h}^{r_0}\sqrt{-V(r)}/f(r)dr$  against $\mu$ at $\Delta=2$ at 10 different temperatures. 
}
\label{bump}
\end{center}
\end{figure}

Next, we would like to comment on the large $\mu$ limit of the charged \ren entropies. When $\mu$ becomes larger than any other scale in the problem, the critical temperature should be proportional to  $\mu$. One should note however, that the ratio $T/\mu$ remains small even in the scaling regime, meaning that the black hole continues to stay close to extremality, an observation also noted in the flatly sliced AdS charged black holes \cite{Hartnoll:2008kx}. This is indeed the case as can be seen in Fig. \ref{q=0}.

It is interesting to compare the critical temperature\footnote{We will actually compare $n_c=1/2\pi T_c$.} of the charged hyperbolic black hole with that of its flat counterpart -- the flat holographic superconductor at finite (physical) chemical potential $\mu$. 
%with the critical temperature at which the charged \ren entropies become discontinuous. 
We find a perfect match for large $\mu$; this is to be expected, since when $\mu/R\gg 1/L$ the curvature of the horizon is small compared to the scale set by the chemical potential. This can be seen in Fig. \ref{q=0} as well. We note that, although the gravitational computations for flat and hyperbolic black holes are quite similar, these quantities have very different CFT interpretations.  In particular, we  see that one can extract information about physical phase transitions at finite chemical potentials and temperatures (which naturally involve higher excited states) solely by considering the entanglement of the ground state.

We should comment on the difference between these two phase transitions from the CFT perspective. 
The superconductor lives on an infinite flat plane and boundary effects do not play any important role.
However, boundary effects at the entangling surface are crucial in the phase transitions of the neutral \ren entropies, as explained in \cite{sachdevON}. 
%This is a strong qualitative distinction between the two type of phase transitions.
In the phase transitions of the charged \ren entropies,
there is a qualitative difference between the neutral and the charged scalar operators. 
As in the case of the neutral \ren entropies, neutral operators can be hosted on the entangling surface without breaking the $U(1)$ symmetry.
Below the critical temperature, these localized operators induce the phase transitions.
There is another instability, which is caused by the entanglement chemical potential term coupling to the entire subsystem.
As in the case of the superconductors, this instability causes the scalar to condense as well\footnote{Neutral scalars may condense if the conformal dimension is small enough}.
These two effects may in principle compete or amplify one another.
On the other hand, charged operators cannot be hosted on the entangling surface unless the $U(1)$ symmetry is spontaneously broken. 
Therefore, it is the entanglement chemical potential effect that causes the phase transition.
This could explain the qualitative difference in the phase transitions between the $q=0$ case and the $q\neq 0$ case (as shown in Fig.\ref{fixedDelta}). 
This would be interesting to investigate from the field theory point of view.%, though we leave this for future work.

Let us comment also on the $n=1$ limit. When $n=1$ and $\mu=0$ the conical defect operator disappears and we obtain the entanglement entropy. 
%, or in other words, the defect operator defining the \ren entropy is withdrawn. 
On the other hand, if $\mu$ is non-vanishing, then even as $n\to1$ a defect operator -- the  Wilson line -- remains. 
So a phase transition could occur even at $n\le1$.
One might worry that a phase transition precisely at $n=1$ would make it impossible to compute entanglement entropies as a limit of the charged \ren entropies.
However, since ${\pd S_{n}\over \pd n}$ is still continuous (only ${\pd^2 S_{n}\over \pd n^2}$ is discontinuous) the \ren entropies are still smooth enough to give unique and well-defined entanglement entropies as $n\to1$.

We note also that the Wilson line couples to the global current,  so at first sight one might expect it not to effect uncharged operators.  We have seen, however, that this is not the case.  While uncharged operators do not directly couple to the Wilson line, they still experience its presence indirectly via couplings to other charged operators. This is manifested holographically by the fact that even neutral scalar fields in the bulk can detect  changes in $\mu$ at fixed temperature indirectly, via the $\mu$ dependence of the metric. 

%In this light therefore, it is reassuring to see 
%becomes a generic feature naturally induced by the defect operator. 

One can also extract  information about the largest eigenvalue of the charged reduced density matrix (\ref{eq:rho0x}) by considering the limit $n\to\infty$.
From the definition of the \ren entropy (\ref{charen}), one can compute the eigenvalues of the charged reduced density matrix (\ref{eq:rho0x}) once we know the
\ren entropy as a function of the \ren parameter $n$;
\bea
\exp((1-n)S_n)=\int_{0}^{\lambda_1}d\lambda~d(\lambda)\lambda^n
\eea
where $\lambda$ is the eigenvalue and $d(\lambda)$ is the spectral density. $\lambda_1$ is the largest eigenvalue. In general,  $d(\lambda)$ contains delta functions. In fact, if the \ren entropy decays polynomially as $n\to \infty,$ one can show that the spectral function take the following form
\bea
d(\lambda)=\delta(\lambda-\lambda_1)h_1(\lambda)+\Theta(\lambda-\lambda_1)h_2(\lambda),
\eea
where $h_1$ and $h_2$ are some functions of $\lambda$ and $\Theta(\lambda-\lambda_1)$ is the Heaviside step function.
There is a simple relation between the largest eigenvalue $\lambda_1$ and the $n\to\infty$ limit of the \ren entropy (called the min-entropy)
\bea
S_{\infty}=-\log \lambda_1.
\eea
Below the critical temperature (when $n>n_c$) the scalar field acquires a non-zero expectation value. These hairy black holes have smaller thermal entropy than that of non-hairy black holes of the same temperature
and chemical potential.
Since the \ren entropy is given by the integral of the thermal entropy (\ref{Rnyi-therm}), this means that the min-entropy $S_{\infty}$ 
is always smaller than that of Einstein-Maxwell black hole
\bea
S_{\infty}(\text{Einstein-Maxwell})>S_{\infty}(\text{Einstein-Maxwell-Scalar})
\eea
Moreover, the critical temperature increases as one decreases the conformal dimension $\Delta$. Therefore, as observed in \cite{Belin:2013dva},
\bea
{dS_{\infty}(\mu,\Delta)\over d\Delta}>0.
\label{ren-dim-ineq}
\eea
The main difference between the neutral case \cite{Belin:2013dva} and the charged case is that  (\ref{ren-dim-ineq}) is a strict inequality even 
in the case of $n_c\le 1$,
while the neutral case is not.  Therefore $S_{\infty}(\mu,\Delta)$ is a monotonic function of $\Delta$.
The entanglement chemical potential dependence of the min-entropy is
\bea
{dS_{\infty}(q\neq0, \Delta)\over d\mu} <0. 
\eea
%On the other hand, although there is a phase transition between $T=T_0$ and $T=\infty$ for $n_c<1$, the $n\to0$ limit of the \ren entropy $S_0$ has a UV divergence and one cannot use it to characterize vacuum states unless properly regularized.

We  close by recalling that the chemical potential in hyperbolic space can be interpreted as the insertion
of a background Wilson line.  The insertion of the Wilson line in the imaginary time direction has opposite orientation when
viewed from region $A$ as opposed to its complement $B$. At the same time, the ground state satisfies $S_n(A)= S_n(B)$, for all $n$, since the reduced density matrices of $A$ and $B$ have the same
eigenvalues.
Thus %he ground state also satisfies
\be
S_n(\mu, A ) = S_n(-\mu, B).
\ee
In a theory with charge conjugation invariance, this would additionally imply that
$S_n(\mu,A)= S_n(-\mu,A)$, i.e. that the charged \ren entropy is an even function of $\mu$.
This is, in particular, clearly true in the case of holographic dual of  Einstein-Maxwell theory.

\section*{Acknowledgments}
We are grateful to O. Dias, M. Headrick, A. Lawrence, R. Myers and A. Nicolis for useful conversations. 
SM acknowledges YITP, Riken, and KEK for hospitality.
This research was supported by the National Science and Engineering Research Council of Canada.

\end{document}